\documentclass[%
 reprint,
 superscriptaddress,
 amsmath,amssymb,
 aps,
 prl,
]{revtex4-2}
\usepackage{xcolor}
\usepackage{lipsum} 
\newcommand{\expe}[1]{\left\langle #1 \right\rangle}

\usepackage{appendix}
\usepackage{braket}
\usepackage{float} 
\usepackage{chemformula}
\usepackage{graphicx}
\usepackage{dcolumn}
\usepackage{bm}
\usepackage{setspace}
\usepackage{braket}
\usepackage{graphicx}
\usepackage{dcolumn}
\usepackage{bm}
\usepackage[colorlinks=true, citecolor=blue, linkcolor=blue, urlcolor=blue]{hyperref}

\makeatletter
\newcommand*{\addFileDependency}[1]{
  \typeout{(#1)}
  \@addtofilelist{#1}
  \IfFileExists{#1}{}{\typeout{No file #1.}}
}
\makeatother

\setcitestyle{super}

\DeclareMathOperator{\sech}{sech}
\DeclareMathOperator{\csch}{csch}

\begin{document}

\preprint{APS/123-QED}
\title{Metastable Photo-Induced Superconductivity far above $T_{\textrm{c}}$}

\newcommand{\affiliationHavard}{
Lyman Laboratory, Department of Physics, Harvard University, Cambridge, MA 02138, USA
}

\newcommand{\affiliationDBSK}{Gwangju Institute of Science and Technology, 123 Cheomdangwagi-ro, Buk-gu, Kwangju, South Korea}

\newcommand{\affiliationRWTH}{
Institut f\"ur Theorie der Statistischen Physik, RWTH Aachen University and JARA-Fundamentals of Future Information Technology, 52056 Aachen, Germany
}
\newcommand{\affiliationMPSD}{
Max Planck Institute for the Structure and Dynamics of Matter,
Center for Free-Electron Laser Science (CFEL),
Luruper Chaussee 149, 22761 Hamburg, Germany
}

\newcommand{\affiliationBremen}{
Institute for Theoretical Physics and Bremen Center for Computational Materials Science,
University of Bremen, 28359 Bremen, Germany
}

\newcommand{\affiliationBristol}{
H H Wills Physics Laboratory, University of Bristol, Bristol BS8 1TL, United
Kingdom 
}

\newcommand{\affiliationETH}{
Institute for Theoretical Physics, ETH Z\"urich, 8093 Z\"urich, Switzerland
}

\newcommand{\affiliationOxford}{
Department	of	Physics,	Clarendon	Laboratory,	University	of	Oxford,	United	Kingdom
}

\author{Sambuddha Chattopadhyay}
\affiliation{\affiliationHavard}

\author{Christian J. Eckhardt}
\affiliation{\affiliationMPSD}
\affiliation{\affiliationRWTH}

\author{Dante M.~Kennes}
\affiliation{\affiliationRWTH}
\affiliation{\affiliationMPSD}

\author{Michael~A. Sentef}
\affiliation{\affiliationBremen}
\affiliation{\affiliationBristol}
\affiliation{\affiliationMPSD}

\author{Dongbin Shin}
\affiliation{\affiliationMPSD}
\affiliation{\affiliationDBSK}
\date{\today}

\author{Angel Rubio}
\affiliation{\affiliationMPSD}
\date{\today}

\author{Andrea Cavalleri}
\affiliation{\affiliationMPSD}
\affiliation{\affiliationOxford}\date{\today}

\author{Eugene A. Demler}
\affiliation{\affiliationETH}

\author{Marios H. Michael}
\affiliation{\affiliationMPSD}
\date{\today}

\begin{abstract}
Inspired by the striking discovery of metastable superconductivity in $\mathrm{K}_3\mathrm{C}_{60}$ at 100K, far above $T_{\textrm{c}}=20K$, we discuss possible mechanisms for long-lived, photo-induced superconductivity. Starting from a model of optically-driven Raman phonons coupled to inter-band electronic transitions, we develop a microscopic mechanism for photo-controlling the pairing interaction. Leveraging this mechanism, we first investigate long-lived superconductivity arising from the thermodynamic metastable trapping of the driven phonon. We then propose an alternative route, where the superconducting gap created by an optical drive leads to a dynamical bottleneck in the equilibration of quasi-particles. We conclude by discussing implications of both scenarios for experiments that can be used to discriminate between them. Our work provides falsifiable explanations for the nanosecond-scale photo-induced superconductivity found in $\mathrm{K}_3\mathrm{C}_{60}$, while simultaneously offering a theoretical basis for exploring metastable superconductivity in other quantum materials. 
\end{abstract}


\maketitle

\begin{figure}[t]
  \centering \includegraphics[width=0.48
\textwidth]{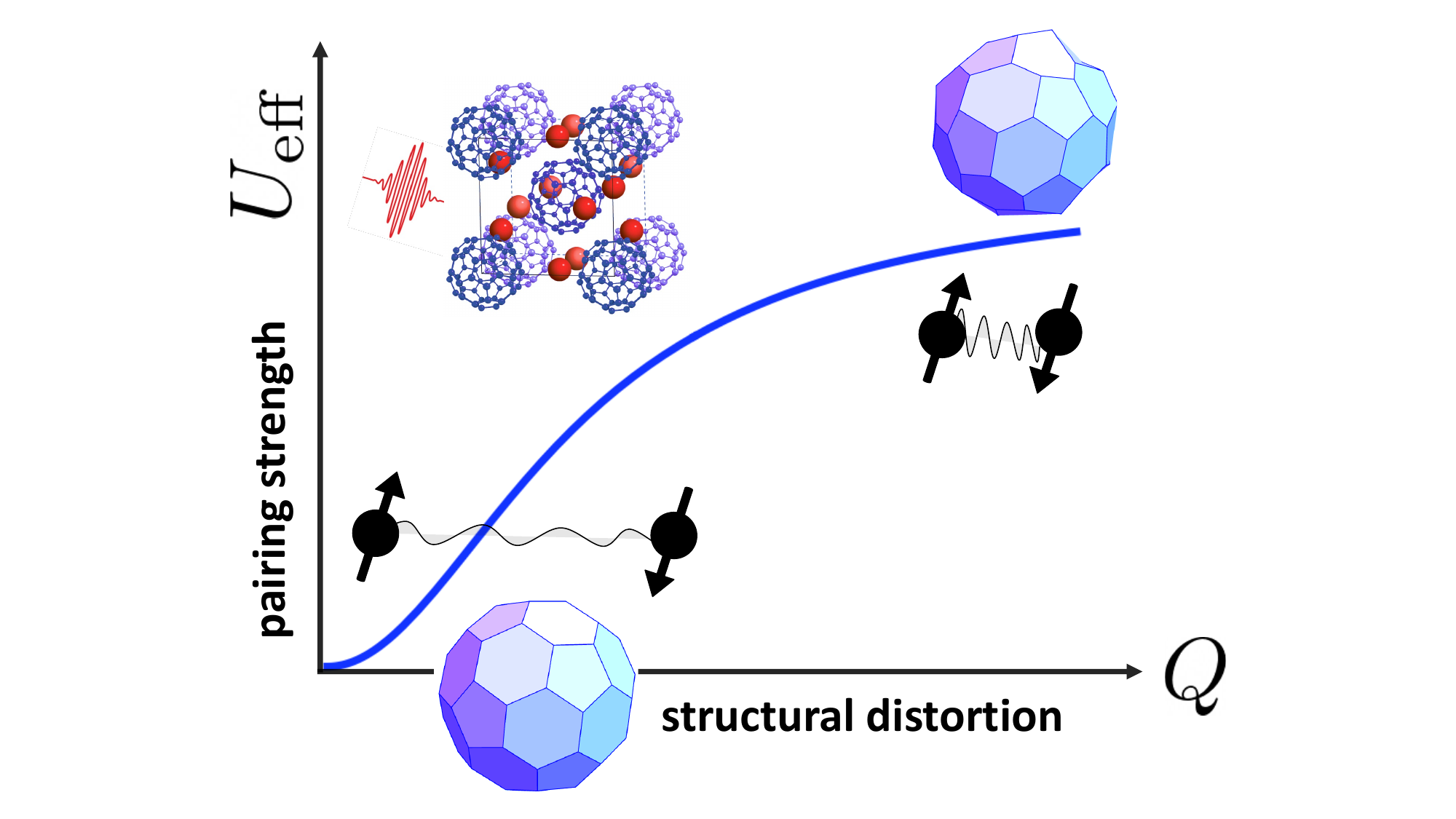}
  \caption{\textbf{Photo-Control of Superconductivity} Schematic plot of the attractive pairing strength between electrons as a function of the photo-induced structural distortion: Inducing local structural distortions (e.g., distorting the on-ball Hg(3) mode in $\ch{K3C60}$) via laser driving provides a route towards controlling superconductivity in molecular solids with light.}
  \label{fig:sketch}
\end{figure}


\section{Introduction}
Breakthroughs in the structural control of matter using intense, infrared (IR) light have enabled the exploration of quantum order outside of the confines of thermal equilibrium \cite{basov_towards_2017,de_la_torre_colloquium_2021}. From photo-induced ferroelectricity in \ch{SrTiO3}\cite{Nova_19,li_terahertz_2019} to laser driven charge density wave ordering in \ch{LaTe3}\cite{Kogar_20,Pavel_20} to optically-stabilized ferromagnetism in \ch{YTiO3} \cite{disaFerromagnetism}  to transient, light-induced superconductivity in layered cuprates \cite{vonHoegen_22,taherian2024,Michael_20,Michael_22}, a striking array of experiments have traced a path towards the photo-control of complex, quantum materials, a focal project for modern condensed matter. Among the most captivating puzzles in this field is the metastable superconductivity uncovered in \ch{K3C60}, far above $T_{c}=20K$ \cite{Mitrano_16,Budden_21,Ed_23}. In the \ch{K3C60} experiments, signatures of photo-induced superconductivity at $T=100$K were observed to persist over 10 nanoseconds \textit{after driving}, 1,000 times longer than any microscopic time-scale in the experiment. The staggering longevity of this response can be further appreciated when contrasted with the lifetime of the \textit{transient} superconductivity observed in, e.g., driven cuprates\cite{vonHoegen_22,taherian2024}, where optical superconducting signatures dissipate within the picosecond scale ring-down time of the resonantly driven apical oxygen phonon. Conceptualizing the \ch{K3C60} experiments thus poses a double-headed challenge: it demands both a microscopic mechanism for photo-induced superconductivity---a topic that has invited intense theoretical attention\cite{kennes_transient_2017, komnik_bcs_2016, nava_cooling_2018, knap_dynamical_2016, raines_enhancement_2015, coulthard_enhancement_2017, kim_enhancing_2016, murakami_nonequilibrium_2017, sentef_theory_2016, Michael_20, denny_proposed_2015, okamoto_theory_2016, dolgirev_periodic_2022, babadi_theory_2017, dasari_transient_2018, mazza_nonequilibrium_2017, dai_superconducting-like_2021,Zhiyuan20}---and an explanation for its metastability.

Galvanized by the experiments performed in \ch{K3C60}, we take on both questions. Within an experimentally motivated minimal model we first describe a microscopic mechanism for non-equilibrium superconductivity arising from the photo-displacement of a local Raman phonon coupled to the inter-band transition between two narrow electronic bands. We then lay out two distinct paradigms for non-equilibrium superconductivity that persists long after driving and provide emergent life-times for the superconducting state. We first compute the adiabatic free energy landscape of our model as a function of the induced structural distortion, uncovering thermodynamically metastable superconductivity far above $T_{\textrm{c}}$. We then advance a dynamical route to long-lived superconductivity---we call this \textit{quasi-particle trapping}. We argue that photo-inducing a large superconducting gap during driving can lead to bottlenecks in the equilibration of quasi-particles; the slow equilibration of quasi-particles then enables the formation of a non-thermal superconducting gap. We find that both pictures may provide inroads towards conceptualizing metastable photo-induced superconductivity in \ch{K3C60} and other molecular superconductors.

\begin{figure*}[t]
  \centering \includegraphics[width=0.96
\textwidth]{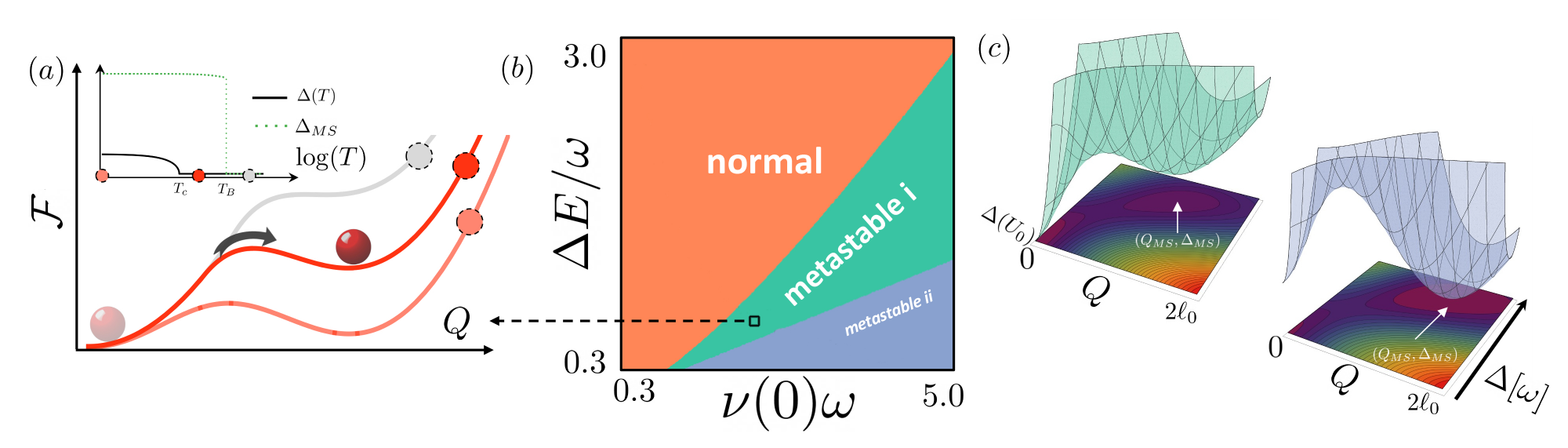}
  \caption{\textbf{Thermodynamically Metastable Superconductivity} Metastability diagram (b) for metastable superconductivity at fixed electron-phonon coupling $\frac{g \ell_0}{\omega}=0.6$ and (second order) $T_{\textrm{c}} = 20 K$ as a function of the level splitting $\Delta E$ and density of states at the Fermi level $\nu(0)$. Distinct regimes of metastability are presented: metastable I (green) and metastable II (blue), with DFT parameters for \ch{K3C60} indicated. Schematic free energy landscape for metastable I (a), depicted across three temperatures---$T=0$ (light red), $T>T_{\textrm{c}}$ (dark red), $T>T_{\textrm{B}}$ (grey). (Inset) Comparison between the gap in the metastable state, $\Delta_{MS}$, and the gap in equilibrium $\Delta(T)$. (c) Free energy landscapes at $T \ll T_{\textrm{c}}$ corresponding to metastable I (left) and metastable II (right) as a function of $\Delta$ and $Q$, in units of $\omega$ and $\ell_0 = \frac{1}{\sqrt{m \omega}}$. In metastable I, The undistorted state is the global free energy minimum: in metastable II, the state with a larger gap due to distortion, is the ground state.}
  \label{fig:metastablesc}
\end{figure*}

\section{Results}

\textit{Model for photo-control pairing---}Inspired by aspects relevant to superconductivity in alkali-doped fullerides---strong, local electron-phonon coupling and narrow bands---we begin with a minimal model that couples dispersionless optical Raman phonons to the local inter-orbital transition between two electronic states. The electrons are allowed to weakly tunnel across sites on a three-dimensional lattice. Specifically, our Hamiltonian $H$ is given by 
\begin{equation}
    H = H_{0} + H_{\textrm{int}} + H_{\textrm{K}},
\label{eq:BasicModelFull}
\end{equation}
where $H_{0} = -\frac{\Delta E}{2} \sum_{i, \sigma}  (c_{i, \sigma} ^ \dag c_{i, \sigma}-d_{i, \sigma} ^ \dag d_{i, \sigma}) + \sum_{i} \frac{P_{i}^2}{2 M }+ \frac{M \omega^2}{2} Q_{i}^2$, the electron-phonon interaction is $H_{\textrm{int}} = g \sum_{i, \sigma} Q_{i} (c_{i, \sigma} ^ \dag d_{i, \sigma}+ d_{i, \sigma} ^ \dag c_{i, \sigma})$, and  the (weak) nearest neighbor hopping Hamiltonian is  $H_{\textrm{K}}=-t \sum_{\langle i, j \rangle, \sigma} (c_{i, \sigma}^\dag c_{j, \sigma} + d_{i, \sigma}^\dag d_{j, \sigma} + h.c.)$. Here $c^\dag_{i, \sigma} (d^\dag_{i, \sigma})$ creates a $c$ (d)-electron with spin $\sigma \in \{\uparrow, \downarrow\}$ at site $i$; $P_i$ and $Q_i$ are the momentum and position coordinates for a phonon of frequency $\omega$ and mass $M$ at site $i$; $\Delta E$ is the splitting between levels $c$ and $d$; $g$ is the electron-phonon coupling. We concern ourselves with a system in which the $c$-band is partially filled at equilibrium.

We consider what happens when the mean local Raman coordinate $\expe{Q}$ is homogeneously distorted by driving the system with intense, off-resonant laser light. For orientation, in the experiments performed on \ch{K3C60}\cite{Budden_21}, the solid is driven at $41~\textrm{THz}$, whereas relevant, strongly coupled Raman phonons are found below $22~\textrm{THz}$. Furthermore, as they are inversion symmetric, Raman modes are driven by the laser non-linearly through the Hamiltonian, $H_{R} \propto E^2(t) Q$, where $E(t)$ is the electric field of the laser light\cite{Gunnarsson_04, Forst_11}. Non-linear, off-resonant driving rectifies the average position of the Raman mode during driving, displacing $\expe{Q}$ and hybridizing the upper and lower bands. To build intuition for this hybridization, we examine the limit of $g \expe{Q} \ll \Delta E$. Shifting $\expe{Q}$ facilitates virtual, inter-orbital transitions between occupied $c$ and unoccupied $d$ levels. A pair of $c$ electrons at the same site can energetically benefit from this virtual tunneling and form a local, singlet pair. As the tunneling is enhanced by increasing $\expe{Q}$, shifting $\expe{Q}$ photo-enhances local pair formation. We rigorize this intuition by deriving in the Methods, for instantaneous $\expe{Q}$, an attractive Hubbard Hamiltonian for effective lower-band $f$ electrons:
\begin{equation}
        H_{\textrm{\textrm{eff}}}= \sum_{k, \sigma} \xi_k  f_{k, \sigma}^\dag f_{k, \sigma}-\frac{U}{N}\sum_{\substack{k, k' \\ |\xi_k|, |\xi_k'|<\omega}}f_{k, \uparrow}^\dag f_{-k, \downarrow}^\dag f_{k', \uparrow} f_{-k', \downarrow},
\label{eq:effectiveHamiltonian}
\end{equation}
where $\xi_k = \epsilon_k-\mu$ and $\epsilon_k$ is the free electron energy dispersion. For a fixed value of $\expe{Q}$, we find that the induced attraction is given by:
\begin{equation}
    U =U_0+\frac{g^2}{M \omega^2} \frac{4 g^2 \expe{Q}^2}{\Delta E_{\textrm{eff}}^2},
    \label{eq:Att}
\end{equation}
where $\Delta E_{\textrm{\textrm{eff}}} = \sqrt{\Delta E^2 + 4 g^2 \expe{Q}^2}$ and $U_0$ is the attractive interaction arising from extrinsic pairing mechanisms that yield superconductivity for the undistorted material. Photo-distorting the strongly coupled Raman mode thus allows us to dynamically enhance the pairing strength. 

\textit{Photo-Induced Metastability---} We first investigate the feasibility of the conventional picture of a long-lived superconducting state accessed by optically driving the material into a hidden, thermodynamically metastable phase \cite{de_la_torre_colloquium_2021} within the scope of our microscopic mechanism. To do so, we examine the free energy of the combined electron-Raman phonon system as a function of $\expe{Q}$ (hereafter $Q$) which captures the free energetic trade-off between the benefit of a displacively induced distortion $Q$---leading to enhanced $U$ and a stronger superconductor---and the elastic cost of the distortion. We ascertain under what conditions the free energy has a non-trivial local minimum---thereby a thermodynamically metastable state---which can be accessed via (sufficiently strong) optical driving. After driving the phonon coordinate into such a metastable trap, a return to equilibrium proceeds through slow thermal nucleation.

Within a mean-field analysis, developed in the Methods, we uncover two types of metastability and, for parameters relevant to $\ch{K3C60}$, provide a phase diagram delimiting their boundaries provided in Fig.~\ref{fig:metastablesc}(b) as a function of the effective density of states at the Fermi energy $\nu(0)$ and the band-splitting $\Delta E$ for a particular distorted phonon with electron-phonon coupling $g$ and frequency $\omega$. The first type, metastability~I (Fig.~\ref{fig:metastablesc}(a)) corresponds to a scenario in which the undistorted state is always the equilibrium state and a metastable state with finite distortion and stronger superconductivity exists between $0<T<T_{\textrm{B}}$. Above $T_{\textrm{B}}$, the free energy monotonically increases with $Q$ and the metastable state becomes untrapped, as shown by the grey landscape in Fig.~\ref{fig:metastablesc}(a). Metastability II, occurs when at, low-temperatures, a distorted state is energetically favorable (see Fig.~\ref{fig:metastablesc}(c)). Here, equilibrium superconductivity exists until a critical temperature $T_{c}$ where a first order phase transition occurs and the superconductivity-inducing distortion is no longer free energetically favorable. In metastability~II, optically switchable superconductivity is possible for $T_{\textrm{c}} < T < T_{\textrm{B}}$. A prominent, equilibrium, experimental signature accompanying metastability~II is the presence of a distortion-driven, first-order superconducting phase-transition. 

In Methods, we detail the mapping between our model and \ch{K3C60}, with an eye towards capturing details around the Fermi Energy, $E_F$. We associate our lower, $c$ band with the lower-lying $t_{1u}$ bands crossing the Fermi energy in \ch{K3C60} and the driven phonon mode to strongly coupled, inter-orbital, on-ball Jahn-Teller Hg phonons. In particular, we identify the $22$ THz Hg(3) phonon as the phonon whose free-energy is most susceptible to hosting a metastable state due to its combination of high-frequency and strong electron-phonon coupling. We note that while the effective parameters translated from DFT (top left, Fig.~\ref{fig:metastablesc}b) lie outside the metastability region, even equilibrium superconductivity in \ch{K3C60} cannot be accounted for by using the DFT band-structure. Instead, strong electronic correlations---arising from the proximity of \ch{K3C60} in equilibrium to Mott insulator phase---can significantly quench the electronic kinetic energy and reshape the band structure, engendering thermodynamic metastability \cite{Capone2002, nomura}.  A distinct signature for metastability is the presence of Jahn-Teller static distortion. For the Hg(3) phonon such a distortion is at the pico-meter scale which, while small, is detectable by ultra-fast, time-resolved diffraction experiments in sufficiently clean systems. 




\textit{Quasi-Particle Trapping---}In the absence of a metastable trap (e.g., as occurs in the normal region in Fig.~\ref{fig:metastablesc}(b)), metastable superconductivity can still arise from the slow equilibration of Bogolyubov quasi-particles, a mechanism we name \textit{quasi-particle trapping}. We begin by considering an undistorted superconductor above $T_{\textrm{c}}$. We then imagine displacing $Q$ to a large $Q^*$, inducing a large pairing attraction $U^*$, where it is held long enough to quasi-equilibrate to a BCS superconducting state with a sizeable gap $\Delta(Q^*,T)$ and low quasi-particle density $n^*$. 

After the drive shuts off, given the lack of a metastable trap, the phonon coordinate relaxes from $Q^*$ to $Q=0$, relaxing the pairing strength from $U^*$ to $U_0$. This relaxation occurs over the ring-down time $\tau_{\textrm{r}}$ of the perturbed Raman mode, the timescale for the driven excitations to damp out. In the $\ch{K3C60}$ experiments $\tau_{\textrm{r}}$ $\sim 10~\textrm{ps}$\cite{Budden_21}. While this is short compared to the lifetime of the metastable superconductor, it is much slower than the inverse gap. Thus while a careful consideration of diabatically excited pairs is necessary if the interaction was quenched from $U^*$ to $U_0$ \cite{grankin}, as the relaxation proceeds adiabatically from the perspective of the superconductor, such effects are negligible. As $Q$ is relaxed back to $0$ and $U$ relaxes back to $U_0$, in order for the electrons to equilibrate, quasi-particles need to be generated out of the condensate via phonon-induced pair-breaking processes. 

\begin{figure*}[t]
  \centering \includegraphics[width=0.98
\textwidth]{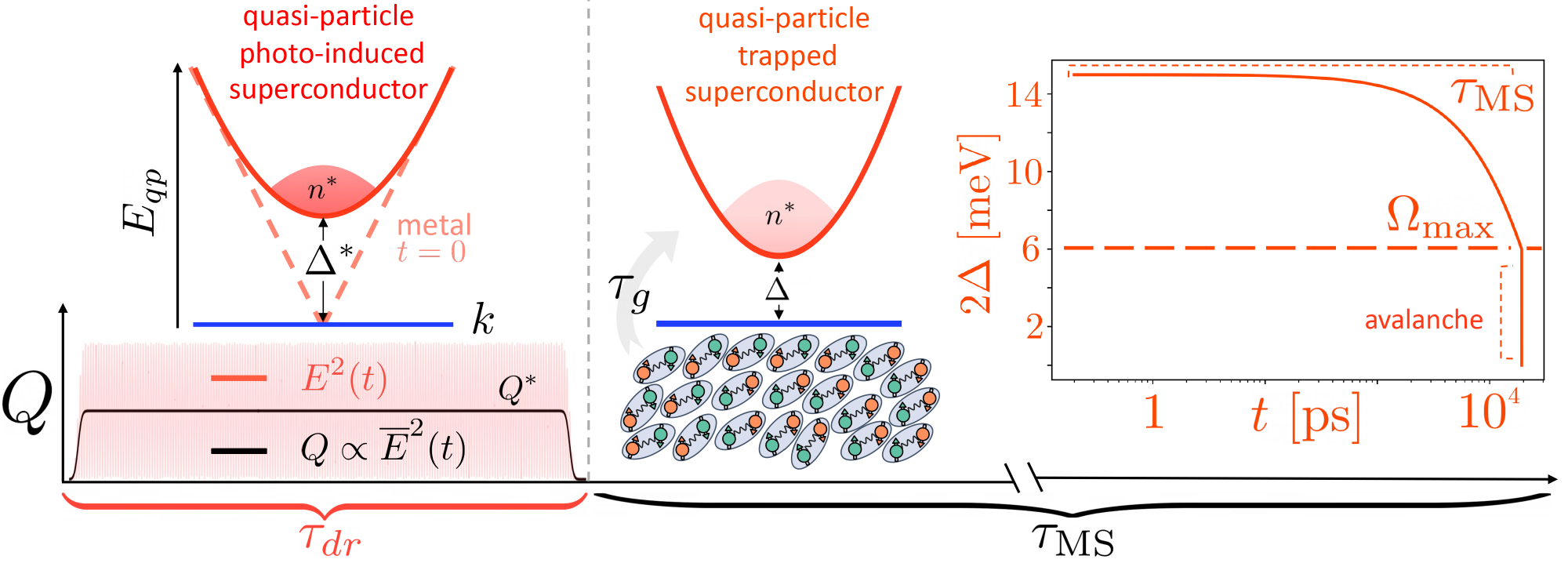}
  \caption{\textbf{Quasi-particle Trapping} Schematic of the ``quasi-particle'' trapping mechanism for long-lived superconductivity. Starting from the metallic phase, displacively driving the phonon coordinate $Q$ from $0$ to $Q^*$ and holding it there photo-induces a superconductor with a gap $\Delta^*$ with a quasi-particle density $n^*$ which is smaller than the thermal quasi-particle density in the metal. After driving for $\tau_{dr}$, the drive is switched off and the phonon relaxes to $Q=0$. However, due to suppressed pair-breaking rates $\tau_g^{-1}$, the quasi-particle density equilibrates slowly, enabling a non-thermal state with gap $\Delta$. (Inset) The dynamics of the photo-induced superconducting gap, persisting out-of-equilibrium for $\tau_{MS}$, until an eventual avalanche collapse.}
  \label{fig:longlivedness}
\end{figure*}

We argue that this process is slow. A large superconducting gap can make scattering processes which conserve quasi-particle number dramatically faster than those which alter the quasi-particle number \cite{RothwarfTaylor,Scalapino}---
While number-conserving quasi-particle scattering involves phonons at arbitrarily low energies, quasi-particle generation, for example, requires the \textit{absorption} of thermally-populated phonons above $2 \Delta$. In the presence of a large gap, the drastic separation of time-scales between the short number-conserving quasi-particle scattering time and much longer  quasi-particle generation and recombination time-scales implies that while the electronic temperature is fixed to the phonon bath temperature due to fast number conserving scattering, the non-equilibrium quasi-particle density evolves slowly. 

Drawing from such considerations, we describe the evolution of the superconducting state after the relaxation of the displaced phonon using an effective, sparse encoding of the quasi-particle spectrum and distribution in terms of two slow variables: the superconducting gap $\Delta(t)$ and a Bogolyubov chemical potential, $\lambda(t)$, respectively. Here $\lambda(t)$ is used to constrain the instantaneous quasi-particle density to a non-thermal value. The joint dynamics of $\Delta(t)$ and $\lambda(t)$, \textit{after} the relaxation of the displaced phonon, can be modelled by solving self-consistent equations for the instantaneous Bogolyubov chemical potential and superconducting gap coupled with an equation for the quasi-particle population dynamics---for details, see the Methods section.

The instantaneous superconducting gap $\Delta(t)$ is given by a modified BCS equation with Debye cutoff $\omega$:
\begin{equation}
     \frac{1}{\nu(0) U_{0}} = \int_0^\omega d\xi \frac{1}{\sqrt{\xi^2 + \Delta(t)^2}} \tanh\left( \frac{\sqrt{\xi^2 + \Delta(t)^2} + \lambda}{2 T} \right),
\label{eq:nonThermalGap}
\end{equation}
which involves a non-equilibrium quasi-particle distribution parameterized by $\lambda$: $n(\omega) =(\mbox{exp}\left(\beta (\omega+ \lambda(t))\right)+1)^{-1}$. Removing thermal quasi-particles (corresponding to $\lambda>0$) by opening up a large gap during driving enables superconductivity above the equilibrium $T_c$, by enabling a greater number of near-degenerate electrons to resonantly participate in pairing. Note that this concept also lies at the core of other mechanisms to enhance superconductivity such as quasi-particle extraction \cite{PhysRevLett.7.274} and the Eliashberg Effect\cite{EliashbergEffect}, the latter developed to conceptualize the enhancement of $T_{\textrm{c}}$ in microwave driven superconductors \cite{Wyatt,Dayem}. We underscore that for large $\lambda$ (small $n$), Eq.~\ref{eq:nonThermalGap} reduces to the BCS equation at $T=0$, a perfect extraction of thermal quasi-particles. Therefore, $\Delta \to \Delta(U_0, T=0)$ as $n \to 0$.

Alongside Eq.~$\ref{eq:nonThermalGap}$, a constraint for the instantaneous quasi-particle density $n(t)$, allows us to fix $\lambda(t)$: 

\begin{equation}
     n(t) = 4 \nu(0)\int_0^\omega d \xi \frac{1}{\mbox{exp}\left(\beta (\sqrt{\xi^2 + \Delta(t)^2} + \lambda(t))\right)+1}.
\label{eq:quasiParticleConstraint}
\end{equation}

Finally, we derive the net dynamics of the quasi-particle density, balancing quasi-particle generation and quasi-particle recombination due to phonon scattering:

\begin{equation}
     \frac{d n}{dt} = \frac{\pi}{2 \hbar} \int_{2 \Delta(t)}^{W} d \Omega  \textrm{ } \alpha^2(\Omega)F(\Omega) \Gamma(\Omega),
\label{eq:qpPopDynamics}
\end{equation}
where the phonon-frequency dependent net quasi-particle generation rate, $\Gamma(\Omega)$, given by 

\begin{multline}
\Gamma(\Omega)  = \Omega^{\frac{1}{2}}\left(\Omega-2\Delta\right)^{-\frac{1}{2}}\csch\left(\frac{\Omega}{2T}\right)  \cr \times \sech\left(\frac{\Delta+\lambda}{2 T}\right)\sech\left(\frac{\Omega-\Delta+\lambda}{2 T}\right) \sinh\left(\frac{\lambda}{T}\right),
\label{eq:pairBreaking}
\end{multline}

is weighted by the Eliashberg function,  $\alpha^2(\Omega)F(\Omega)$, the electron-phonon coupling weighted phonon density of states; where the minimum phonon frequency to break/form Cooper pairs is given by $2 \Delta(t)$ and the maximum phonon frequency is set by the electronic bandwidth $W$, the largest electronic energy transfer possible for dissipative, resonant electron-phonon scattering. As expected, $\lambda>0$---a dearth of quasi-particles compared to equilibrium---implies a net generation of quasi-particles, as is required for equilibration. 

A brief comment on the initial conditions of our post-relaxation dynamics is in order. First, note that we operate under the following ordering of time-scales: $\frac{1}{\Delta(0)} \ll \tau_{r} \ll \tau_g$, where $\tau_g^{-1} = \frac{dn}{dt}|_{t=0}$. This implies that during the relaxation of the phonon, the quasi-particle density is frozen, therefore $n(0) = n^*$, the quasi-equilibrium, quasi-particle density attained upon driving. However, as $U$ has relaxed to the equilibrium attraction $U_0$, the initial gap $\Delta(0)$ is given by the solution of Eq~\ref{eq:nonThermalGap}, with $U_0$ and $\lambda(n^*)$. $\Delta(0) \sim \mathcal{O}(\Delta (T=0))$ is typically much smaller than $\Delta^*$.

In a solid with an unstructured phonon continuum, the transient gap will generically collapse rapidly as quasi-particle generation decreases the gap, increasing the quasi-particle generation rate leading to an ``avalanche" feedback loop. \ch{K3C60} and other molecular solid superconductors, however, possess a crucial structural motif in their phonon spectra: A continuum of low-frequency inter-molecular modes (e.g. acoustic, librational modes) and relatively sharp, high-frequency, local intra-molecular vibrational modes, separated by a sizeable phonon gap. If after the driven phonon relaxes, $2 \Delta(0)$ lies inside the phonon gap, the avalanche collapse of the gap is prevented and the gap decrease instead very slowly due to quasi-particle generation from thermally suppressed high-frequency intra-molecular modes. This slow evolution of the gap persists until it reaches the upper bound of the inter-molecular continuum $\Omega_{\textrm{max}}$, after which the gap avalanche collapses. 

Remarkably, this optimistic scenario is viable in \ch{K3C60}. While the precise nature of electron-phonon coupling in the material is intensely debated \cite{Gunnarsson_04}, an incontrovertible aspect of the phonon spectrum is the existence of a large phonon gap between inter- and intra-molecular modes: Neutron scattering experiments \cite{NeutronScattering} find a precipitous drop in the phonon density of states starting at $6$ meV, persisting to $28$ meV. Moreover, an estimate of the Eliashberg function computed from an inversion of reflectance data \cite{opticalProperties, NeutronScattering} implies that $\alpha^2(\Omega)F(\Omega)$ is negligible in the vicinity of $10-33$ meV. On the other hand, experimentally measured values of $2  \Delta(T=0K)$---the largest initial $2 \Delta(0)$, achievable via ``perfect quasi-particle extraction"---is believed to lie within this gap \cite{TunnelingSpectroscopy, MonolayerK3C60}. Furthermore, a conservative estimate of the gap extracted from optical conductivity in the photo-induced metastable state is consistent with $2 \Delta \sim 14~\textrm{meV} > \Omega_{\textrm{max}} \sim 6-10~\textrm{meV}$ \cite{Mitrano_16}.

\begin{figure}[t]
  \centering \includegraphics[width=0.48
\textwidth]{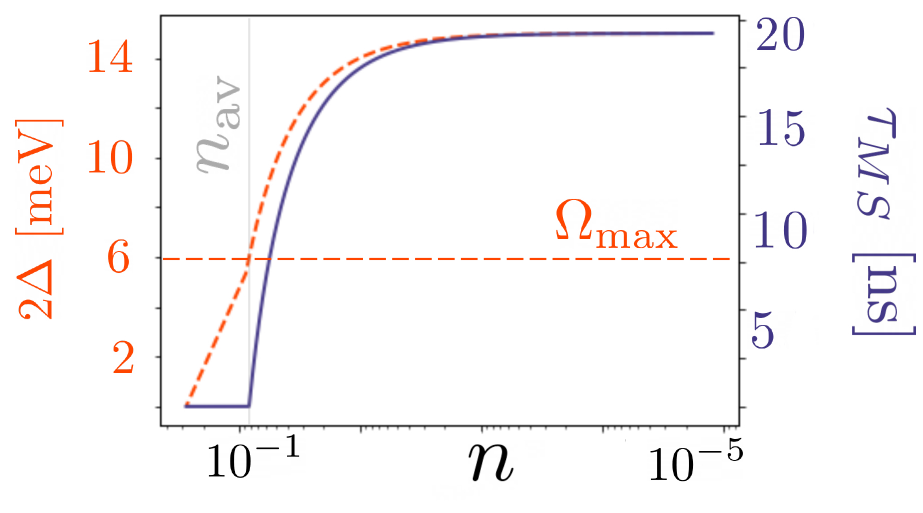}
  \caption{\textbf{Metastable lifetime \& avalanche threshold} 
  The superconducting excitation gap $2 \Delta$ (orange) and the corresponding lifetime of the quasi-particle trapped state $\tau_{\textrm{MS}}$ (navy) as a function of the density of quasi-particles $n \equiv n(0)$ after driving.  
  Reducing the quasi-particle density below a threshold $n_{av}$ avoids the rapid, avalanche collapse of the gap. Experimentally plausible parameters for $\ch{K3C60}$ reproduce the nanosecond scale life-time observed in experiments for sufficiently strong driving.} \label{fig:avalanche}
\end{figure}

If $2 \Delta(0)$  lies in the phonon gap, a steady build-up of quasi-particles occurs at a roughly fixed rate, $\tau_{\textrm{g}}^{-1}$, set by the molecular vibrational modes above $2 \Delta$, until the avalanche density $n_\textrm{av}$---the quasi-particle density corresponding to a transient gap of $\Omega_\textrm{max}/2$---is reached. Thus, an approximate lifetime of the superconducting state is given by $\tau_\textrm{MS} \approx  n_\textrm{av} \tau_\textrm{g}$, where for a set of sharp, vibrational modes $\Omega_\nu \gg T, \Delta(0)$, $\tau_\textrm{g}$ is:

\begin{equation}
    \tau_\textrm{g}^{-1} = \sum_{\nu} g_{\nu}^2 \textrm{  }\nu_{\textrm{SC}}\left(\Omega_{\nu}-\Delta(0)\right)\exp{\left(-\frac{\Omega_{\nu}}{T}\right)},
\label{eq:gen}
\end{equation}

with $g_\nu$, the electron-phonon coupling for mode $\nu$, and $\nu_{\textrm{SC}}$, the superconducting density of states. As discussed in detail in the Methods, the simultaneous suppression of charge fluctuations and electronic bandwidth due to strong repulsive interactions\cite{Capone2002, Gunnarsson_04,nomura} along with weak Hg(1) coupling\cite{bellK3C60} in \ch{K3C60} would permit a metastable lifetime of nearly $20~\textrm{ns}$ at $100 \textrm{K}$. Furthermore, as shown in Fig~\ref{fig:avalanche}, depleting the quasi-particle density during driving to a value below but at the order of the threshold density $n_\textrm{av}$ is sufficient to reach the $10~\textrm{ns}$ scale life-time observed in the experiment. 

\section{Discussion}
In our work we present a microscopic mechanism for inducing non-equilibrium superconductivity by photo-distorting inter-band Jahn-Teller modes. We subsequently offer two alternate paradigms for metastability: first, a conventional picture of optically-switching into a hidden thermodynamically metastable state and second, a dynamical paradigm for metastable superconductivity resulting from the slow thermalization of quasi-particles. In our quantitative considerations, we find while both mechanisms are not beyond physical plausibility, quasi-particle trapping can yield the staggering experimentally observed lifetime within assumptions that are supported by optimistic material parameters obtained from the literature. 

These two pictures can be distinguished by performing Rothwarf-Taylor style quasi-particle injection experiments in the metastable state\cite{RothwarfTaylor}. Here one would expect that the quasi-particle trapped state should be significantly more susceptible to gap collapse than the thermodynamically metastable superconductor upon injection of quasi-particles: For the latter, once a density of $n_{\textrm{av}}$ quasi-particles is reached, avalanche collapse is inevitable. Another key distinguishing feature between the two pictures is the size of the metastable gap. Whereas in the quasi-particle trapping scenario the metastable gap is bounded by the zero temperature gap, a thermodynamically metastable state typically has a much larger superconducting gap than in equilibrium (see, Fig.~\ref{fig:metastablesc}a inset as an example). Accurate experimental extraction of the non-equilibrium superconducting gap---avoiding contamination from other spectral features (e.g. the polaronic peak starting at $10~\textrm{meV}$ occluding precise optical extraction in \ch{K3C60})---may also be used to discriminate between quasi-particle trapping and thermodynamic metastability mechanisms.


This work opens up several directions that are ripe for future investigation. Given the central, albeit subtle, role that strong repulsive interactions play, subsequent work should rigorously examine the interplay between strong Mott repulsion and Jahn-Teller pairing in the driven setting, seeking to resolve if the cooperation between the two that is implicated in equilibrium \cite{Capone2002, Gunnarsson_04} extends to the photo-induced case. 

Future work addressing metastable superconductivity without invoking a thermodynamically hidden metastable state should also aim to resolve the (drastic) asymmetry between the longevity of the state and the time it takes to prepare it. In the quasi-particle trapping scenario articulated above, we claimed by postulate that, through laser-driving, a quasi-equilibrium state with a large gap $\Delta^*$ and significantly reduced quasi-particle density $n^*$ could be realized. Naively, thermalizing towards a state with a large induced attractive interaction would require significant quasi-particle recombination which, like quasi-particle generation, is very slow compared to number conserving scattering for superconductors with a large gap. A first-order concern may be that given that recombination is slow, preparing a metastable quasi-particle trapped superconductor necessitates driving for long-times.

Such anxieties can be partially addressed. The asymmetry between the preparation time and lifetime of the state in this case, however, could arise from a distinct asymmetry between quasi-particle recombination and quasi-particle generation rates: While the latter involves phonon emission the former involves absorption. Thus, while generation is bottlenecked by a lack of thermal-phonons, recombination is slow because of a lack of thermally populated quasi-particles \cite{RothwarfTaylor}. Notice however, that when preparing the quasi-particle trapped state, one starts with an \textit{excess} of quasi-particles, implying that recombination is fast at the beginning of driving \footnote{Within a minimal quasi-particle kinetics model, the optimal ratio between lifetime and preparation time at $T=100K$ is $30$, consistent with the longest drive durations reported in the experiment \cite{Budden_21}}. 

Beyond quasi-particle kinetics, one may speculate that for shorter drive durations the dynamical instabilities implicated in the formation of the superconducting gap play a crucial role. Indeed, theoretical analysis of instantaneous quenches of the Feschbach mediated pairing interaction in ultra-cold fermions suggests that the gap formation dynamics are governed by the near-integrable dynamics of the superconductor, not quasi-particle kinetics \cite{Barankov_2004}. In particular, in the cold atoms setting, for a rapid quench from a free Fermi gas to a superconductor with a large gap $\Delta^*$, the order parameter grows exponentially with rate $\gamma \sim \Delta^*$ to $\Delta \sim \mathcal{O}(\Delta^*)$, followed by damped solitonic oscillations for $T \ll \Delta^*$, e.g. temperatures germane to quasi-particle trapping. Naively extrapolating these results to the regime where the drive duration $\Delta^* \tau_{\textrm{dr}} \gg 1$---i.e. taking $\frac{d\Delta}{dt} = \gamma(t) \sim \Delta(t)$ for $\Delta \tau_{\textrm{dr}} \gtrsim 1$---implies that a gap on the order of $\Delta^*$ should form over a time-scale shorter than $\tau_{\textrm{dr}}$. Such considerations inspire future investigations of the precise interplay between dynamical instabilities and quasi-particle kinetics in solid-state superconductors at finite temperatures, studies which should illuminate how the non-equilibrium superconducting state can be robustly formed from a metal at incredibly short times---empirically, as short as $1.5 \textrm{ps}$.

Finally, we advertise that recent experiments\cite{Ed_23} uncover not only a resonant enhancement of photo-induced superconductivity for driving between $11-15~\textrm{THz}$---enabling superconductivity at significantly lower intensities---but also, tantalizingly, non-equilibrium superconductivity at room temperature. While these experiments do not interrogate metastability explicitly, they reveal that signatures of superconductivity do not vary over at least $50 \textrm{ps}$. Intriguingly, the lifetime of a quasi-particle trapped state at room temperature---using the same parameters used to attain a life-time of $20~ \textrm{ns}$ at $100 K$---is $0.4~\textrm{ns}$, far exceeding this experimental lower bound. On the microscopic side, the resonant features around $11 \textrm{THz}$ are suggestive of photo-induced superconductivity arising from the \textit{resonant} driving---vis-a-vis, displacive excitation---of the $22~\textrm{THz}$ Hg(3) phonon. The non-equilibrium dynamics of a driven phonon coupled to inter-band transitions was recently considered, albeit in a different experimental context, by some of the authors\cite{eckhardt2023theory} \textcolor{red}, showing the enticing opportunity of resonantly enhanced transient superconductivity. Interrogating whether such effects are indeed metastable demands understanding the competition between resonantly enhanced superconductivity and amplified electronic dissipation, a daunting challenge to be grappled with in future investigations.


.




\section{Methods}
\textit{$\expe{Q}$-dependent superconductivity}---
In this section, using a local electron-phonon coupling model, we derive a $\expe{Q}$ dependent BCS-type $s$-wave pairing interaction by performing a judiciously chosen rotation and a disentangling Schrieffer-Wolff transformation. We begin by diagonalizing the solely electronic part of the Hamiltonian of Eq.~\ref{eq:BasicModelFull} in momentum space:
\begin{widetext}
   \begin{equation}
    H = \sum_k \xi_k \left( c_k^\dag c_k + d_k^\dag d_k \right) -  \frac{\Delta E}{2} \sum_k\left( c_k^\dag c_k - d_k^\dag d_k \right) + g \sum_i Q_i \left(c^\dag_i d_i + d_i^\dag c_i \right) + \sum_i \frac{Pi_i^2 }{2 M }+ \left( \frac{M \omega^2 Q_i^2 }{2}  \right).
\end{equation}
\end{widetext}

Here, again for ease of readability,  $c^\dag_{i, \sigma} (d^\dag_{i, \sigma})$ creates a $c$ (d)-electron with spin $\sigma \in \{\uparrow, \downarrow\}$ at site $i$; $P_i$ and $Q_i$ are the momentum and position coordinates for dispersionless phonons of frequency $\omega$ and mass $M$ at site $i$; $\Delta E$ is the splitting between levels $c$ and $d$; $g$ is the local electron-phonon coupling. We concern ourselves with a system in which the $c$-band is partially filled at equilibrium.

In the presence of a fixed distortion $\expe{Q}$, we re-write $Q_i = \expe{Q} + \tilde{Q_i}$, where $\tilde{Q_i}$ are the fluctuations of the phonon coordinate. Our aim is to diagonalize the (non-interacting) electronic Hamiltonian as a function of the structural distortion $\expe{Q}$, thereby accounting for the crucial, non-perturbative enhancement of level-hybridization between the $c$ and $d$ levels arising from increasing $\expe{Q}$. To carry this out, we perform the following rotation:
\begin{equation}
  \begin{pmatrix} f_k \\ g_k  \end{pmatrix} = \begin{pmatrix} \cos(\theta) && \sin(\theta) \\
  -\sin(\theta) && \cos(\theta) \end{pmatrix}  \begin{pmatrix} c_k \\ d_k \end{pmatrix} ,
\end{equation}

where $\theta$ is given by $\tan(2\theta) =  \frac{2 g \expe{Q} }{\Delta E}$. Upon rotating to  $f$ and $g$ electrons, our Hamiltonian becomes: 

\begin{widetext}
   \begin{equation}
    H = \sum_k \xi_k \left( f_k^\dag f_k + g_k^\dag g_k \right) -  \frac{\Delta E_{\textrm{eff}}}{2} \sum_k\left( f_k^\dag f_k - g_k^\dag g_k \right) + 2 g \sum_q \tilde{Q}_q\left( \cos(\theta) S^x_q + \sin(\theta)S^z_q \right) + \sum_i \left(  \frac{P_i^2 }{2 M } + \frac{M \omega^2 Q_i^2 }{2}  \right) ,
\end{equation}
\end{widetext}

where $S_q^{z} = \frac{1}{2} \left( f^\dag_{k+q} f_k - g^\dag_{k+q} g_k \right)$, $S_q^{x} = \frac{1}{2} \left( f^\dag_{k+q} g_k + g^\dag_{k+q} f_k \right)$, and $\Delta E_{\textrm{eff}} = \sqrt{\Delta E^2 +  ( 2 g \expe{Q} ) ^2}$. For $\Delta E_{\textrm{eff}} \gg \omega$, we can restrict our attention to the effective lower $f$-band and write our effective Hamiltonian as: 

\begin{equation}
\begin{aligned}
   H_{\textrm{eff}} &= \sum_k \xi_k f_k^\dag f_k + g \sin(\theta) \sum_{q } Q_{-q} f^\dag_{k+q} f_k  \\
   &+\sum_i \left( \frac{P_i^2 }{2 M } + \frac{M \omega^2 Q_i^2 }{2}  \right) ,
\end{aligned}
\end{equation}

We can now decouple the electronic and phonon degrees of freedom to lowest order in $g$ by using the following Schrieffer-Wolff unitary:
\begin{equation}
    U = e^{ i \sum_q \left( \alpha_q P_q + \beta_q Q_{-q}\right) f^\dag_{k+q} f_k},
\end{equation}

where $\alpha_q, \beta_q$ satisfy: 
\begin{align}
     \beta_q =& M \alpha_q \left( \xi_{k+q} - \xi_k \right) ,\\
    \alpha_q =& - \frac{g \sin(\theta)}{M\left( \omega^2 - \left(\xi_{k+q} - \xi_k \right)^2 \right)}.
\end{align}

Upon decoupling electrons and phonons, we are left with an interacting electronic Hamiltonian with:
\begin{equation}
    H_{\rm int} = - \sum_{k,k',q} \frac{g^2 \sin^2(\theta)}{2 M \left( \omega^2 - \left(\xi_{k+q} - \xi_k \right)^2 \right)} c_{k+q,\sigma'}^\dag c_{k'-q,\sigma}^\dag c_{k', \sigma } c_{k,\sigma'}.
\end{equation}

Similar to how retardation is captured in BCS theory, we approximate $H_{\textrm{int}}$ above with an instantaneous interaction with $\expe{Q}$ dependent interacting $U = \frac{g^2 \sin^2(\theta (\expe{Q})}{2 M \omega^2 }$ and an energy cut-off at $\omega$. All told, this yields our desired attractive interaction Eq.~\ref{eq:Att}:

\begin{equation}
    \delta U(\expe{Q})=\frac{g^2}{M \omega^2} \frac{4 g^2 \expe{Q}^2}{\Delta E_{\textrm{eff}}^2},
\end{equation}

\textit{Mean-Field Metastability}---We now explore the simplest quasi-equilibrium picture for long-lived, light-induced superconductivity: optically driving the material into a hidden, thermodynamically metastable phase \cite{de_la_torre_colloquium_2021}. We study the driven, dissipative dynamics of the mean phonon coordinate $Q$ in an adiabatic free energy landscape that weights the free-energy benefit of inducing a displacive shift $Q$---leading to enhanced $U$ and a stronger superconductor---and the elastic cost of such a distortion. We work within a finite-temperature BCS ansatz for the electrons. Note that while this at first may be seem ill suited to describe superconductivity for putatively non-BCS superconductor $\ch{K3C60}$, following the suggestion proffered by Capone et. al. \cite{Capone2002}, such an approach is both tractable and sensible for considering the superconductivity of the reduced quasi-particle weight, free-fermion like quasi-particles. Accordingly, when considering $\ch{K3C60}$ specifically, we keep in mind quasi-particle spectrum obtained from DMFT. 

Using a finite-temperature BCS ansatz for the electrons, we model the evolution of $Q$ as:

\begin{equation}
\bigg(\partial_t^2 + \gamma \partial_t + \frac{1}{M}\frac{\partial \mathcal{F}_{\textrm{SC}}}{\partial Q}+\omega^2\bigg)Q = zE^2(t)+\eta(t),
\end{equation}

where $zE^2(t)$ is the effective displacive acceleration arising from laser driving; where $\eta(t)$ is fluctuation-dissipation obeying white noise with $\langle \eta(t) \eta(t') \rangle = 2 \frac{\gamma}{M} T \delta(t-t')$; and where the BCS superconducting free-energy density $\mathcal{F}_{SC}$ is given by:
\begin{widetext}
\begin{equation}
    \mathcal{F}_{\textrm{SC}}= \frac{\Delta^2}{U} - 2 \nu(0) \int_0^{\omega} d \xi \left( \sqrt{\xi^2 + \Delta^2 } - \xi \right) - 4 \nu(0) T \int_0^{\omega} d \xi \ln \left( 1 + e^{-\beta \sqrt{\xi^2 + \Delta^2}}\right).     
\end{equation}
\end{widetext}

Here $\nu(0)$ is the effective density of states per spin per site at the Fermi-level and $\Delta(Q)$ is obtained by solving the BCS self-consistency equation for $U(Q)$.

We start with a system without a shift ($Q=0$), in the normal state, at finite temperature $T>T_{\textrm{c}}$. We assume that the drive is sufficiently strong to traverse any free-energy barrier between the global free-energy minimum at $Q=0$ and the putative, superconducting metastable state at $Q_{\textrm{MS}} \neq 0$. After driving the phonon coordinate into such a metastable trap, a return to equilibrium through macroscopic nucleation due to thermal fluctuations is, canonically, slow. Our goal is thus to ascertain when the free energy landscape has a non-trivial local minimum. A sufficient (and from numerical observations, necessary) condition for the free energy landscape to host a metastable state is that $T=0$ energy landscape should have a non-trivial local minimum. For $U_0=0$, a metastable state exists if:
\begin{equation}
    Q_{\textrm{c}}^2 = \frac{\Delta E \mbox{ }\omega}{\sqrt{2} g^2} \Delta(Q_c), 
\end{equation}
has two positive solutions $Q_{\textrm{B}}$ and $Q_{\textrm{MS}}$, corresponding to a local maximum and minimum in the energy landscape, respectively. Metastability persists until $T_{B}$ where the \textit{free} energy landscape no longer has non-trivial critical points and the metastable state becomes untrapped, as shown by the grey landscapes in Fig.~\ref{fig:metastablesc}(a).

We reiterate the distinction between the two types of metastability that we find. The first type, metastability~I, corresponds to the scenario in which $E(Q_{\textrm{MS}}) > E(0)$, where $E(Q) = \frac{1}{2}M \omega^2 Q^2 - \omega^2 \nu(0) \mathrm{coth}\bigg(\frac{1}{\nu(0) U(Q)}\bigg)$. In this scenario, the distortionless, normal state is always the equilibrium state and metastability occurs between $0<T<T_{\textrm{B}}$. The alternate scenario, metastability II, occurs when at $T=0$ a distorted ground state is energetically favorable. Equilibrium superconductivity exists until a critical temperature $T_{c}$ where a \textit{first} order phase transition occurs and the superconductivity-inducing distortion is no longer free energetically favorable. In metastability~II, optically switched superconductivity is possible for $T_{\textrm{c}} < T < T_{\textrm{B}}$. In passing, we note that while our approach near the local maximum is within BCS, the precise position of the local minima, thereby the quantitative boundary between metastability I \& II, may require a treatment beyond BCS to capture the effects of strong fluctuations.

\textit{Superconductivity with fixed quasi-particle number 
}---In this section we provide a derivation of the steady state equations Eqs.~\ref{eq:nonThermalGap} \& \ref{eq:quasiParticleConstraint} within the BCS mean field approximation. We note that similar considerations were asserted previously by Parmenter\cite{PhysRevLett.7.274} in the case of quasi-particle extraction. We begin by reiterating that (number conserving) scattering between quasi-particles and thermally occupied acoustic phonons is much faster than number non-conserving scattering events needed to generate quasi-particles. While the latter is emphasized throughout our work as what enables a long-lived non-equilibrium gap, the former suggests that the effective temperature that governs the quasi-particle distribution should coincide with the temperature of the phonon bath---we do not invoke a two-temperature model. Thus, to find our steady state quasi-particle distribution, it suffices to solve for the gap $\Delta$ and Bogolyubov chemical potential $\lambda$ self-consistently while minimizing an effective thermal free energy at temperature $T$. Concretely, our quasi-particle number constrained, $\Delta$-dependent free energy can be expressed as: 

\begin{equation}
\begin{aligned}
    F(\Delta) &= \mbox{Tr} \{ \rho H (\Delta)\} + T \mbox{Tr}\{ \rho \log \rho \} + z \left( \mbox{Tr} \{\rho\} - 1 \right) \\ 
    &+ \lambda \left( \mbox{Tr} \{ \rho \sum_{k,\sigma} \gamma^\dag_{k,\sigma} (\Delta) \gamma_{k,\sigma} (\Delta) \} - n\right),
\end{aligned}
\label{eq:FreeEnergyAppendix}
\end{equation}

where the first two terms correspond to the usual BCS decoupled free energy at a given temperature and superconducting gap $\Delta$, the fixed quasi-particle constraint is enforced by minimizing with respect to a Lagrange multiplier $\lambda$, and the final term enforces that the density matrix has trace 1, by minimizing with respect to $z$.  The operators $\gamma_{k,\sigma} (\Delta)$ ($\gamma_{k,\sigma}^\dag (\Delta)$) correspond to the Bogolyubov quasi-particle annihilation (creation) operators defined by canonical Bogolyubov rotations parametrized by $\Delta$: $\gamma_{k,\uparrow} = u_k c_{k, \uparrow}-v_{k} c_{k, \downarrow}^{\dag}$ and $\gamma_{-k,\downarrow}^{\dag} = u_k^* c_{-k, \downarrow}^{\dag}-v_{k}^* c_{k, \uparrow}$, where $|u_k|^2 = \frac{1}{2}\left(1+\frac{\xi_k}{\sqrt{\xi_k^2+\Delta^2}}\right)$ and $|v_k|^2+|u_k|^2=1$. We note now that the operator portion of the quasi-particle number constraint can be subsumed into the Hamiltonian, which within BCS mean-field, is diagonal in $\gamma_{k,\sigma} (\Delta)$. Thus, one readily interprets the Lagrange multiplier as a chemical potential for the Bogolyubov quasi-particles. 

For completeness, by factorizing our density matrix into different momentum $k$-dependent and psuedo-spin $\sigma$-dependent contributions,
\begin{equation}
    \rho = \Pi_{k,\sigma} \rho_{k,\sigma}, \mbox{\qquad with \qquad} \mbox{Tr}\{ \rho_{k,\sigma} \} = 1. 
\end{equation}
The $\rho_{k,\sigma}$'s are found by minimizing $F$:
\begin{equation}
\begin{aligned}
    \frac{\partial F}{\partial \rho_{k,\sigma}} =& E_k \expe{\gamma^\dag_{k,\sigma} \gamma_{k,\sigma}}  \\
    &+ T \log(\rho_{k,\sigma}) + T  + \lambda \expe{\gamma^\dag_{k,\sigma} \gamma_{k,\sigma}} + z = 0,\\
    \Rightarrow \rho =& \frac{e^{ - \beta \sum_{k, \sigma}\left( E_k + \lambda \right) \gamma^\dag_{k,\sigma} \gamma_{k,\sigma}}}{Z_k}, 
\end{aligned}
\end{equation}
where $\lambda$ and $\Delta$ are determined self-consistently as:
\begin{equation}
    n = \sum_{k,\sigma} \expe{\gamma^\dag_{k,\sigma}\gamma_{k,\sigma}}  = \sum_{k,\sigma}  \frac{1}{\mbox{exp}\left(\beta (\sqrt{\xi_k^2 + \Delta^2} + \lambda)\right)+1}
\end{equation}

\begin{widetext}
\begin{equation}
    \Delta = \frac{U}{N}\sum_{\substack{k \\ |\xi_k|<\omega}} \expe{c_{-k, \downarrow}c_{k, \uparrow}}  = \frac{U}{N} \sum_{\substack{k \\ |\xi_k|<\omega}}  u_k^*v_k (1-2\expe{\gamma_{k,\uparrow}^\dag \gamma_{k,\uparrow}})=\frac{U}{N}\sum_{\substack{k \\ |\xi_k|<\omega}} \frac{\Delta}{2 \sqrt{\xi_k^2 + \Delta^2}} \tanh{\left(\beta (\sqrt{\xi_k^2 + \Delta^2} + \lambda)\right)} 
\end{equation}
\end{widetext}

\textit{Relaxation Time-Scales: Pair-Breaking \& Quasi-Particle Recombination}---In this section, we derive a set of characteristic time-scales relevant to the equilibration of Bogolyubov quasi-particles coupled to a thermal bath of phonons, supporting Eqs.~\ref{eq:qpPopDynamics} \textrm{-} \ref{eq:gen} in the main-text. We note that the bath we consider contains weakly coupled, low frequency lattice / intermolecular modes up to a cutoff $\Omega_{\textrm{max}}$ and sharp, high-frequency optical modes $\Omega_\nu$. The details of the former are not particularly relevant as they set the precise dynamics of the avalanche. For the former, we take the interaction to be of the local electron phonon type, reflecting the coupling between electrons with molecular vibrational modes of a molecular solid. 

We first derive the dynamics due to the low-frequency phonons. We introduce a weakly coupled bath of low-energy, phonons, with electron-phonon interactions given by:

\begin{equation}
    H_{\textrm{bath}} =\sum_{p, p',\lambda, \sigma} g_{p, p', \lambda} (a_{q, \lambda}+a_{-q, \lambda}^\dag) c^\dag_{p, s} c_{p, \sigma}.
\label{eq:bath}
\end{equation}

where $\lambda$ runs over the polarizations of the phonons, $g_{p, p', \lambda}$ yields a momentum and polarization dependent electron-phonon coupling, $q = p-p'$---considering non-Umklapp processes---and $\sigma \in \{\uparrow, \downarrow\}$. Leveraging the assumption that electrons are coupled weakly to the low-lying phonons, we estimate the quasi-particle-phonon scattering times perturbatively using Fermi's Golden Rule. We delineate between three different equilibration rates: (1) $\Gamma_{\textrm{s}} = \frac{1}{\tau_s}$, the rate of scattering between Bogolyubov quasi-particles by the acoustic phonons \textit{without changing the total number of quasi-particles}, (2) 
$\Gamma_{\textrm{r}} = \frac{1}{\tau_r}$ the rate at which quasi-particles re-combine into Cooper-pairs, and an additional rate: $(3)$ $\Gamma_{\textrm{g}} = \frac{1}{\tau_{\textrm{g}}}$, the rate at which quasi-particles are generated due to pair breaking due to phonons. 

We compute the Fermi's golden rule rate by using $H_{\textrm{bath}}$ as the perturbation and picking an initial state $\ket{I}$ sampled from a density matrix formed by the Kronecker product of a thermal density matrix of phonons at a temperature $T$ and, for the electrons, an arbitrary quasi-particle occupation function\cite{schrieffer_2018}. In thermal equilibrium, this is the Fermi-Dirac distribution; in the quasi-particle trapped steady state, this is given by $f(E) = \frac{1}{\mbox{exp}\left(\beta (E + \lambda)\right)+1}$. Throughout our discussion, we focus on rates where one of the quasi-particles involved is at the gap $\Delta$. We do this because while in principle the rate is frequency-dependent, the density of states of the superconductors is strongly peaked at the gap. Moreover, as we coarse-grain over the time-scale over which the electrons thermalize to the phonon temperature, we are mainly concerned with the net influx rate of quasi-particles generated / recombined. As such, a typical scale which provides the correct dynamics is given by $\Gamma_{\textrm{r / g}}(\Delta)$. 

In the ensuing, we focus first on process (3), the generation of quasi-particles by scattering with bath phonons: the first two processes are discussed in detail in \cite{Scalapino}. We make our discussion concise by computing the generation time for a fixed quasi-particle at the gap energy $\Delta$ and with momentum $k=0$ and spin $\uparrow$. The process whose rate we seek to compute is one which involves the absorption of a phonon at energy $\Omega$ and the generation of two quasi-particles, one at energy $\Delta$, the other at energy $\Omega-\Delta \geq \Delta$. We write one possible final state as $\ket{F}_{q,\lambda} = \gamma^\dag_{k, \uparrow} \gamma^\dag_{q, \downarrow} a_{k-q,\lambda} \ket{I}$. To compute the total scattering rate, we sum over quasi-particle momenta $q$. 

We make progress by noting that terms such $c_{p', \uparrow}^\dag c_{p, \uparrow} + c_{-p, \downarrow}^\dag c_{-p', \downarrow}$ can be written in terms of Bogolyubov quasi-particle operators as: $c_{p', \uparrow}^\dag c_{p, \uparrow} + c_{-p, \downarrow}^\dag c_{-p', \downarrow} = t(p,p')(\gamma_{p, \uparrow}^\dag\gamma_{p',\uparrow}+\gamma_{-p, \downarrow}^\dag\gamma_{-p',\downarrow})+m(p,p')(\gamma_{p', \uparrow}^\dag\gamma_{-p,\downarrow}^\dag-\gamma_{p, \uparrow}\gamma_{-p',\downarrow})$, where $t(p,p')$ and $m(p,p')$ are the standard coherence factors: $m^2(p,p') = \frac{1}{2}(1+\frac{\epsilon_p \epsilon_{p'}+\Delta^2}{E_p E_{p'}})$ and $t^2(p,p') = \frac{1}{2}(1+\frac{\epsilon_p \epsilon_{p'}-\Delta^2}{E_p E_{p'}})$. Using this, we can express the square of our desired matrix elements as: $|\bra{I} H_{bath} \ket{F}_{q,\lambda}|^2 = g_{k-q, \lambda}^2 m^2(k,q) (1-\nu_{q, \downarrow}) (1-\nu_{k, \uparrow}) n_{ph}(k-q, \lambda)$, where $\nu_q$ is the occupation number of a quasi-particle at momentum $q$ and $n_{ph}(k-q, \lambda)$ is the occupation number of a phonon at momentum $k-q$ and polarization $\lambda$ in the initial state $\ket{I}$. Note that while $\ket{I}$ is a pure state, the rate will be averaged over occupation numbers given by the ensemble from which $\ket{I}$ is drawn from, as prescribed above.

Computing the sum over $q$ in the continuum, we introduce the standard Bogolyubov density of states---for a quasi-particle at energy $E$, the DOS goes as $\rho(E) = \frac{E}{\sqrt{E^2-\Delta^2}}$---and phonon spectral density $F(\Omega)$, weighted by the square of the matrix element $\alpha^2(\Omega)$, averaged over the Fermi surface\cite{Scalapino}. We note that while taking such an $\alpha^2(\Omega)F(\Omega)$ is appropriate for interactions with a low-frequency bath, given that the electronic density of states changes dramatically at higher-frequencies, naively using such an $\alpha^2(\Omega)F(\Omega)$ would be inappropriate in this context. 

Using our computation of the matrix element above, we arrive at our result for the rate at which quasi-particles are generated by the scattering of Cooper pairs with low-frequency phonons:

\begin{widetext}
\begin{equation}
\Gamma_{\textrm{g}}(\Delta) = \frac{2 \pi}{\hbar} \int_{2\Delta}^\infty d \Omega \left( \alpha^2(\Omega) F(\Omega) \sqrt{\frac{\Omega}{\Omega-2\Delta}}(1-f(\Omega-\Delta))(1-f(\Delta)) n(\Omega)\right)
\label{eq:generation}
\end{equation}    
\end{widetext}

The scattering time $\tau_s$ which controls thermal equilibration within each quasi-particle band is given by: 

\begin{widetext}
\begin{equation}
\Gamma_{\textrm{s}}(\Delta) = \frac{2 \pi}{\hbar} \int_{0}^\infty d \Omega \left( \alpha^2(\Omega) F(\Omega) \sqrt{\frac{\Omega}{\Omega+2\Delta}}(1-f(\Delta+\Omega))(1-f(\Delta)) n(\Omega)\right),
\end{equation}    
\end{widetext}

Note that $\Gamma_{\textrm{s}}$ can be much larger that $\Gamma_{\textrm{g}}$ as low-frequency acoustic modes can be significantly thermally populated as arbitrarily low-frequency phonons can participate in number-conserving scattering processes. Our arguments for a fast scattering, $\tau_{\textrm{s}}$ time but a very slow generation time, $\tau_{\textrm{g}}$, provide justification for our approximation that a superconducting steady state is reached described by a temperature and a conserved number of quasi-particles. 

Finally, for completeness, we provide the rate equations for quasi-particle recombination:

\begin{widetext}
\begin{equation}
\Gamma_{\textrm{r}}(\Delta) = \frac{2 \pi}{\hbar} \int_{2\Delta}^\infty d \Omega \left( \alpha^2(\Omega) F(\Omega) \sqrt{\frac{\Omega}{\Omega-2\Delta}} f(\Omega - \Delta)f(\Delta) (1 + n(\Omega) ) \right).
\label{eq:recombination}
\end{equation}    
\end{widetext}

Having argued for the large discrepancy between scattering rates that thermalize the Bogolyubov quasi-particles and rates for generating new quasi-particles, we now argue that the asymmetry between quasi-particle recombination during driving and quasi-particle generation after driving implies that our long-lived superconductor can be generated by a short pulse, as seen in experiments\cite{Ed_23, Budden_21}. While generating quasi-particles out of the condensate can be slow due to the absence of thermally excited phonons, recombining a pair of quasi-particles requires emitting a phonon, and is therefore not bottlenecked by the phonon population. The recombination bottleneck in traditional ``Rothwarf-Taylor" quasi-particle injection experiments arises instead from a dearth of thermally excited quasi-particle partners that are needed for recombining the injected quasi-particle\cite{RothwarfTaylor}: In the quasi-particle injection experiments $f(\Omega - \Delta) \sim e^{-\frac{\Omega-\Delta}{T}}$ at low-temperatures. In our context, however, if one rapidly ramps $U$ to $U^*$, the recombination process is not slowed down due to a \textit{lack} of quasi-particles: in this situation, initially there is not a lack but an \textit{excess} of non-thermally distributed quasi-particles that need to be recombined.

Finally, we arrive at the question of how to modify Eq. ~$\ref{eq:generation}$ and Eq.~$\ref{eq:recombination}$ for the case of the local, vibrational modes. For these---obviating for now the difference between inter- and intra-band couplings---we consider a Holstein type model: $H_{\textrm{e-ph}} = \sum_{i, \nu} g_{\nu} Q_i n_i$, summing over modes $\nu$ and sites $i$. Moving to momentum space and performing the same Fermi's Golden Rule analysis as above---suppressing coherence factors as $\Omega_\nu \gg \Delta$---reveals that the proper way to modify $\alpha^2(\Omega)F(\Omega)$ to the high frequency setting is to write $\alpha^2(\Omega)F(\Omega) = \sum_{\nu} g^2 \nu_{\textrm{SC}}(\Omega_\nu-\Delta)\delta(\Omega-\Omega_\nu)$. Considering this rate for $\Omega_\nu \gg \Delta, T$ yields Eq.~$\ref{eq:gen}$. 

In passing, we note that the dissipative processes considered here are single phonon relaxation mechanisms. In principle, at sufficiently high temperatures, multi-phonon processes arising from, e.g. thermally populated acoustic modes, could dominate relaxation due to single phonon optical phonons. A precise accounting for this is complicated by the lack of momentum resolved couplings for acoustic modes. Moreover, it is unclear if they play a significant role in \ch{K3C60} as acoustic modes are believed to not have significant coupling to electrons in the material \cite{Gunnarsson_04}. Nevertheless, a quantitative accounting of the relevance of multi-phonon processes should be developed to study the relaxation of driven superconducting systems.


\textit{Mapping to \ch{K3C60}}---In this section we articulate our microscopic model, describing how it captures essential features of $\ch{K3C60}$. We begin by reviewing the literature relevant to the phonon-pairing based (alternatively, the inverse Hunds)  theory of superconductivity \ch{K3C60} and use it to guide an association between our effective two electronic band, single phonon model and the e.g. three band, five phonon physics of $\ch{K3C60}$. Our goal is to achieve quantitative estimates that dictate how plausible our pictures of longevity are to explaining the metastable superconductivity uncovered in $\ch{K3C60}$. 

We begin by unpacking the predominant theory regarding the origin of high-$T_{\textrm{c}}$ equilibrium, s-wave superconductivity in \ch{K3C60} which implicates an interplay between attraction facilitated by Jahn-Teller Hg modes that are strongly coupled to partially filled $t_{1u}$ bands and strong electronic correlations that arise from the materials proximity to the Mott transition \cite{Gunnarsson_04, Capone2002, nomura}. The latter leads to a dramatic suppression of charge fluctuations and significant renormalization of both the onsite Coulomb repulsion (in the absence of Tomalchev-Anderson renormalization, due to the large values of $\omega$ as compared to $E_f$) and the bandwidth ($W \to ZW$, where the quasi-particle residue, $Z \ll 1$), due to Brinkman-Rice physics \cite{Gunnarsson_04}. The Jahn-Teller modes induce an ``inverse Hund's'' coupling via the dynamical Jahn-Teller effect that provides the attractive pairing glue for local Cooper pairs. Crucially, as the Jahn-Teller interaction is ``traceless'' with respect to the charge degrees of freedom, the suppression of charge fluctuations due to Mott repulsion does not effect the attractive inverse Hund's coupling, as opposed to electron-phonon interactions with Ag modes which couple to the density \cite{Gunnarsson_04}: strong electron-phonon interactions and electronic correlations work cooperatively to lead to a $T_{\textrm{c}} \sim 20K$. 

In light of the centrality of Mott physics to superconductivity in \ch{K3C60}, we carry out the mapping of our minimal model to the electronic physics of \ch{K3C60} \textit{around $E_F$} from an unconventional but incisive vantage point, starting neither from vanilla DFT models of the electronic structure of \ch{K3C60} nor from the molecular limit but instead interpreting DMFT calculations that account explicitly for Mott effects \cite{nomura}. The DMFT modified spectral function near the Fermi energy contains two sharp quasi-particle peaks---dramatically sharper than in DFT---which we associate with two partially filled, narrow $t_{1u}$ bands crossing the Fermi surface and one empty, narrow $t_{1u}$ band, along similar lines to the interpretation of DMFT performed by Capone et. al. in Ref \citep{Capone2002}. As the two peaks in the Mott modified spectral function  overlap far from the Fermi energy, we make contact with \ch{K3C60} by identifying the lower $c$ band in our model with the two half-filled $t_{1u}$ bands crossing the Fermi surface and the upper $d$ band with the empty $t_{1u}$ band. We map $\Delta E$ in our model to the separation of the centers between the two sharp spectral peaks. To summarize, as a crude but comprehensive approximation, we assume that the Mott physics in \ch{K3C60} acts simply to  renormalize the repulsive interaction and electronic spectral features such as the bandwidth and band-splitting---beyond this, the superconductivity is determined by local electron-phonon physics. 

Having made these identifications on the electronic side, we note that the Jahn-Teller Hg phonons couple to the $t_{1u}$ bands with different polarizations providing both inter-band and intra-band transitions \cite{Gunnarsson_04}. Within our effective two-band model, we take polarizations coupling to intra-band transitions to give rise to the equilibrium pairing attraction $U_0$ that balances the inverse Hund's coupling with the Mott renormalized Coulomb interaction to give $T_{\textrm{c}} \sim 20$K, s-wave superconductivity in equilibrium. Inter-band polarizations coupling the half filled bands to the empty bands, give an additional contribution to attraction, $U(Q)$, as we have shown in our model. We note in passing that virtual inter-orbital Cooper tunneling is already believed to play a small contributing role in equilibrium superconductivity \cite{nomura, Rice}, underscoring the interpretation of our microscopics as photo-enhancing a (virtual) Suhl-Kondo effect. To determine whether a metastable superconducting state exists in \ch{K3C60} we take the electron-phonon coupling $g$ for each Hg mode to be consistent with  ab-initio calculations \cite{ePHCoupling, bellK3C60}. We find that the optimal mode for photo-induced metastability in \ch{K3C60}---possessing both strong electron-phonon coupling and high-frequency---is the Hg(3) mode at $22$ THz: the relevant polarization for this mode has $\frac{g \ell_0}{\omega} \approx 0.6$, where $\ell_0$ is the effective oscillator length of the phonon \cite{ePHCoupling, bellK3C60}. Note that while every mode that can be non-linearly rectified, will be, only the modes whose free-energy landscapes have metastable states can be trapped. 

\label{appendix:PairBreaking}
\begin{figure}[t]
  \centering \includegraphics[width=0.48
\textwidth]{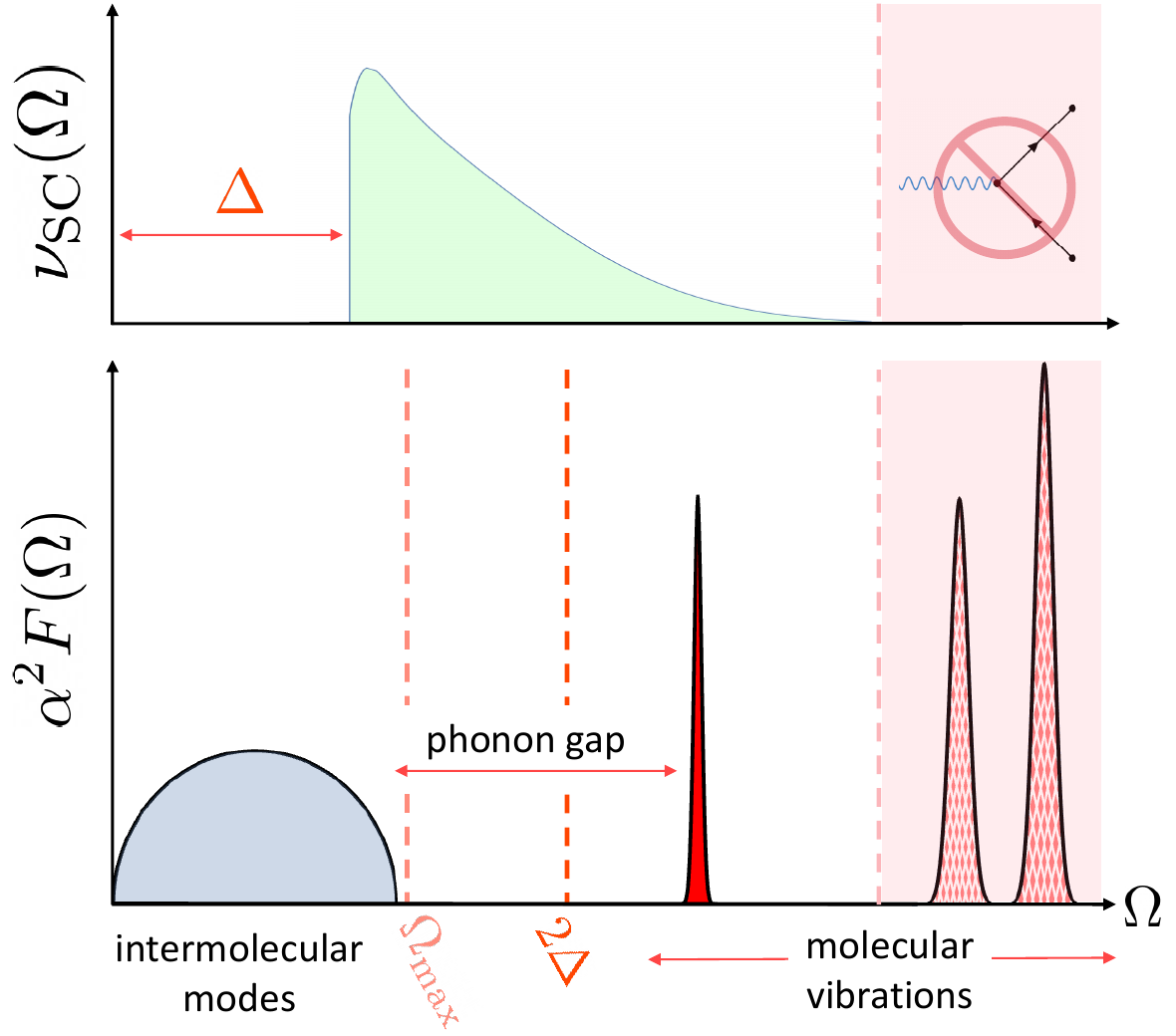}
  \caption{\textbf{Electron, Phonon Density of States in $\ch{K3C60}$} Schematic of superconducting electron (top) and phonon (bottom) density of states in $\ch{K3C60}$. As discussed in detail, the phonon spectrum contains a significant gap, separating lattice inter-molecular modes of the $\ch{K3C60}$ crystal from the on-ball molecular vibrational modes. In the optimal quasi-particle trapping scenario, strong driving pushes $2 \Delta$ into the phonon gap, e.g. above $\Omega_\textrm{max}$. Note that while the on-ball vibrations can provide dissipation for the Cooper pairs, a narrow electronic band can suppress resonant electronic scatters: Above the electronic bandwidth quasi-particle generation is kinematically constrained.}
  \label{fig:phonons}
\end{figure}

\textit{Pair-Breaking: An Interplay of High-Frequency Phonons \& Narrow Bands}---We begin by reminding the reader that the generation rate for quasi-particles for a superconductor whose charge gap ($2 \Delta$) after driving lies in between the intermolecular and intramolecular modes is given by:

\begin{equation}
    \tau_\textrm{g}^{-1} = \sum_{\nu} g_{\nu}^2 \textrm{  }\nu_{\textrm{SC}}\left(\Omega_{\nu}-\Delta(0)\right)\exp{\left(-\frac{\Omega_{\nu}}{T}\right)},
\label{eq:gen_app}
\end{equation}

where $\Omega_\nu$ are the frequencies corresponding to high energy, sharp, local, vibrational phonon modes. In $\ch{K3C60}$, the $Hg(3)$ phonons that are relevant to superconductivity are also the modes that are relevant to dissipation. Moreover, modes that are not of Jahn Teller flavor (e.g. not trace-less in the orbital degrees of freedom) do not couple strongly to the effective delocalized quasi-particles because charge fluctuations are dramatically suppressed due to strong Coulomb repulsion \cite{hanK3C60}. Beyond the presence of the phonon gap and the importance of the Jahn-Teller Hg modes, there is significant controversy regarding the precise quantitative nature of electron-phonon coupling in the material, although Hg(3) and $\textrm{Hg(8)}$ modes are often implicated \cite{Gunnarsson_04}. To attain the nanosecond time-scale at $100 K$ reported in experiments within the quasi-particle trapping paradigm, it is also important to neglect dissipation due to the lowest lying Hg(1) mode at $33 \textrm{meV}$, something which is supported by estimates in $\ch{K3C60}$ for high carbon bond-stretching to bond-bending ratio\cite{bellK3C60}. Thus, we take contributions starting from the Hg(2) phonon at $52 \textrm{meV}$. To obtain a long-lifetime, a interplay of thermal suppression and of the (superconducting) density of states must overpower the strong local electron phonon coupling. The most relevant contribution thus comes from the $52 \textrm{meV}$ mode. Using molecular parameters from \textit{ab initio} calculations\cite{bellK3C60} for the electron phonon coupling, we arrive at a constraint for the effective bandwidth of the electrons (i.e. as is necessary to suppress $\nu_{\textrm{SC}}$). To get the necessary suppression of the superconducting density of states---and approximating the mobile quasi-particle density of states as a Gaussian---we find that a bandwidth of $\sim 60 \textrm{meV}$ is sufficient to produce a maximal $20 \textrm{ns}$ lifetime for the state. Note that while $60 \textrm{meV}$ is small, it is very much in agreement with seminal DMFT results of Capone et. al. \cite{Capone2002}, who find that a quasi-particle residue $Z \sim 0.06$ is necessary for superconductivity due to the a local, phonon induced inverse Hund's coupling in the presence of strong repulsive interactions. From this perspective, a reduction of the $\sim 0.5 \textrm{eV}$ DFT bandwidth by $Z \sim 0.06$ is roughly half as much as is necessary for the long-lifetime. 

\section{Data Availability}

The data presented in this work is produced directly by code which can be made available upon request.

\section{Code Availability}
All codes used to generate the figures are available upon request.

\providecommand{\noopsort}[1]{}\providecommand{\singleletter}[1]{#1}%

\section{Acknowledgements}
We acknowledge stimulating discussions with E. Rowe, G. Jotzu, M. Buzzi, B. Halperin, A. Polkovnikov, S. Gopalakrishnan, A. Georges, M. Devoret, P.A. Lee, M. Eckstein, G. Refael, A. Auerbach, \& A.J. Millis. S.C. is grateful for support from the NSF under Grant No. DGE-1845298 \& for the hospitality of the Max Planck Institute for the Structure and Dynamics of Matter. M.H.M. would like to acknowledge the support from the Alexander von Humboldt Foundation. D.S. was supported by the National Research Foundation of Korea (NRF) grant funded by the Korea government (MSIT) (No. RS-2023-00253716 and RS-2023-00218180). ED acknolwedges support from the ARO grant number $\textrm{W911NF-21-1-0184}$ and the SNSF project $200021-212899$. We also acknowledge support from the European Research Council (ERC-2015-AdG694097), the Cluster of Excellence ``Advanced Imaging of Matter'' (AIM), Grupos Consoldados (IT1453-22), Deutsche Forschungsgemeinschaft (DFG) – SFB-925 – project 170620586, DFG -- Cluster of Excellence Matter and Light for Quantum Computing (ML4Q) EXC 2004/1 -- 390534769 (within the RTG 1995), DFG - 508440990, and the Max Planck-New York City Center for Non-Equilibrium Quantum Phenomena. The Flatiron Institute is a division of the Simons Foundation.

\section*{Author Contribution}
M. H. M. conceived the project together with S.C. and E. A. D.. S.C., M. H. M. and E. A. D. developed the theoretical and analytical framework. S. C., M. H. M. and D. S. analyzed existing experimental and \textit{ab initio} data. S.C. developed and performed all numerical calculations. M. H. M., E. A. D., M. A. S., D. M. K., A. R. and A.C. sponsored and supervised the project. All authors participated in the discussion and interpretation of the results. S.C., C. J. E., M. A. S., E. A. D., A. C., and M. H. M  wrote the manuscript with input from all authors. \\

\section{Competing interests}
The authors declare no competing interests.

\end{document}